\documentclass[a4paper,10pt,fleqn]{article}
\usepackage[dvips,colorlinks,bookmarksopen,bookmarksnumbered,linkcolor=black,citecolor=black,urlcolor=black,breaklinks]{hyperref}
\usepackage{amsmath}
\usepackage{graphicx}
\usepackage{breakurl}
\usepackage[authoryear]{natbib}
\usepackage{times}
\usepackage{mathrsfs}
\usepackage{color}

\begin{document}
\title{Survival analysis for AdVerse events with VarYing follow-up times (SAVVY): Rationale and statistical concept of a meta-analytic study}

\author{Regina Stegherr$^1$, Jan Beyersmann$^1$, Valentine Jehl$^2$,\\
 Kaspar Rufibach$^3$, Friedhelm Leverkus$^4$, Claudia Schmoor$^5$, \\
and Tim Friede$^6$\footnote{Corresponding author: Tim Friede, Institut f\"ur Medizinische Statistik, Universit\"atsmedizin G\"ottingen, Humboldtallee 32, 37073 G\"ottingen, Germany, {\sf{e-mail: tim.friede@med.uni-goettingen.de}}, Phone: +49-551-39-4991, Fax: +49-551-39-4995} \hspace{.1cm} on behalf of the SAVVY project group \\ \\
\normalsize $^1$ Institut f\"ur Statistik, Universit\"at Ulm, Ulm, Germany \\
\normalsize $^2$ Novartis Pharma AG, Novartis Pharma AG, Basel, Switzerland \\
\normalsize $^3$ Methods, Collaboration, and Outreach Group (MRO), Department of Biostatistics, \\
\normalsize Hoffmann-La Roche Ltd., Basel, Switzerland \\
\normalsize $^4$ Pfizer, Berlin, Germany \\
\normalsize $^5$ Clinical Trials Unit, Faculty of Medicine and Medical Center,\\
\normalsize University of Freiburg, Freiburg im Breisgau, Germany \\
\normalsize $^6$ Institut f\"ur Medizinische Statistik, Universit\"atsmedizin G\"ottingen, G\"ottingen, Germany}

\date{19 November 2019}

\maketitle 

\begin{abstract}
  The assessment of safety is an important aspect of the evaluation of new
  therapies in clinical trials, with analyses of adverse events being an
  essential part of this. Standard methods for the analysis of adverse events
  such as the incidence proportion, i.e. the number of patients with a
  specific adverse event out of all patients in the treatment groups, do not
  account for both varying follow-up times and competing risks. Alternative
  approaches such as the Aalen-Johansen estimator of the cumulative incidence
  function have been suggested. Theoretical arguments and numerical
  evaluations support the application of these more advanced methodology, but
  as yet there is to our knowledge only insufficient empirical evidence
  whether these methods would lead to different conclusions in
  safety evaluations. The Survival analysis for AdVerse events with VarYing
  follow-up times (SAVVY) project strives to close this gap in evidence by
  conducting a meta-analytical study to assess the impact of the methodology
  on the conclusion of the safety assessment empirically. Here we present the
  rationale and statistical concept of the empirical study conducted as part
  of the SAVVY project. The statistical methods are presented in unified
  notation and examples of their implementation in R and SAS are provided. 
\end{abstract}

\section{Introduction} 
Time-to-event or survival endpoints are commonly
encountered in clinical trials, and some literature searches have found that
survival analysis is the most common advanced statistical technique in medical
research \citep{Hort:Switz:2005,sato2017statistical}. Censoring is arguably
the major reason to use survival methodology, since statistical inference that
does not account for censoring will, in general, be biased. Primary efficacy
outcomes in time-to-event studies are often comprehensive
outcomes such as overall or progression-free survival
\citep{schumacher2016competing}, and Kaplan-Meier curves are commonly used to
estimate survival probabilities, while the log-rank test and/or the Cox model
are used to compare treatment groups \citep{gosho_2018}. Here, a
`comprehensive outcome' is a time-to-event outcome that every
patient experiences, possibly after study closure which would then result in
an administratively censored observation. There are, however,
concerns that many published Kaplan-Meier curves are subject to a so-called
`competing risk bias' \citep{van2016competing,schumacher2016competing}. Such
bias arises for non-comprehensive time-to-event outcomes, and Kaplan-Meier
will then overestimate cumulative event probabilities.

Safety evaluation is an essential aspect of clinical trials
  beyond the efficacy evaluation of new therapies \citep{yang2018} with
  primary focus on quantifying the incidence of adverse events
(AEs). But the use of sophisticated survival methodology in
practice does not translate to AE analyses, and, arguably, the major
workhorses to quantify AE incidence are the incidence proportion, i.e., the
number of patients with an observed AE (of a certain type) divided by group
size, and the (exposure adjusted) incidence density, which divides by
cumulative patient-time at risk. This gap, using time-to-event
  methodology for efficacy analyses but not for AE
analyses, has been criticized by a number of authors for some
time already, see \citep{oneill_1987,allignol2016,bender_2016, unkel2018}. The
issue is that the incidence proportion estimates the probability that a
patient experiences an AE \textit{and} that it is observed before censoring
which is less than the absolute AE risk, i.e., the probability of experiencing
the AE. Closely related to this issue are varying follow-up times which
further complicate using incidence proportions, see, in particular,
\cite{bender_2016}.

  The incidence density, on the other hand, accounts for both censoring and
  varying follow-up times by considering patient-time at risk in the
  denominator rather than the simpler number of patients. The incidence
  density is not an estimator of absolute AE risk, but rather of the AE hazard
  under a constant hazard assumption. This rather restrictive parametric
  assumption has been repeatedly criticized
  \citep{Krae:even:2009,bender_2019}. The Kaplan-Meier curve has been
  considered as a non-parametric alternative on the probability scale
  \citep{siddiqui2009statistical,crowe2009}, but is subject to the aforementioned
  competing risk bias. This would also be true for its parametric counterpart,
  when translating the incidence density onto the probability scale.  

  The nature of competing risk bias is that one minus a Kaplan-Meier
  curve approximates an empirical distribution function, where the
  approximation is due to incompletely observed data as a consequence of
  censoring. Empirical distribution functions eventually reach 100\%. For
  estimating absolute AE risk by the Kaplan-Meier method, the consequence is that
  one implicitly assumes that every patient would experience the AE under
  consideration eventually. However, this does not hold true for patients who die before
  the AE, and the absolute AE risk is consequently overestimated. Here, `death
  before AE' acts as a competing event (or `competing risk') and must be
  accounted for in the statistical analysis. For instance, Allignol et al.\
  \citep{allignol2016} demonstrate in a real data analysis that a simple
  parametric AE analysis using incidence densities but accounting for competing
  events may outperform a Kaplan-Meier analysis.

  So, the concern is that quantifying absolute AE risk may be
  either underestimated (incidence proportions) or overestimated
  (Kaplan-Meier, incidence densities). The issues are mainly censoring,
  varying follow-up times, competing events and, in the case of incidence
  densities, a possibly too restrictive parametric model. Here, a
  non-parametric benchmark method is provided by the Aalen-Johansen estimator
  \citep{Aale:Joha:an:1978} which generalizes the Kaplan-Meier estimator to
  multiple event types, see \cite{jbcs19} for a recent textbook treatment in
  the context of AE. The magnitude of such bias in practice on, e.g., AE
  frequency categories will, however, depend on the frequency of competing
  events, the magnitude of censoring or the difference in follow-up.  In terms
  of between-group comparisons, the impact of, say, dividing two metrics that both underestimate or both
  overestimate is also unclear.

  The SAVVY (Survival analysis for AdVerse events with VarYing follow-up
  times) project strives to close this gap in evidence by conducting a
  meta-analytical study. In this project, participating sponsor companies and organizations (in the following called sponsors) select randomized controlled clinical trials of special interest, particularly those with varying follow-up times between patients and possibly between treatment arms. This includes studies in different therapeutic areas. Within studies, one or more AEs of interest will be selected.
  Here we present the rationale and statistical
  concept. The statistical methods are presented in unified notation and the
  implementation in R and SAS is described adding some interest to this work
  beyond presenting concepts of the main study. While basic methodological
  considerations on AE analyses, censoring, varying follow-up times and
  competing risks have been discussed elsewhere, see, e.g., \cite{unkel2018}
  and \cite{jbcs19}, this paper offers additional detailed insights by
  considering practically relevant questions such as which kind of event
  should be viewed as competing and which one as a standard censoring
  event. For instance, \cite{unkel2018} briefly touch upon the question
  whether diagnosed progression is a competing event or rather a censoring
  event, and if the latter, whether such censoring is informative, and,
  finally, what the impact on AE analyses is. Here, Section~\ref{sec:CE}
  offers further guidance, also distinguishing between a ``hard'' competing
  event definition and a more encompassing one. Furthermore, an important
  meta-analytical issue is that estimates are typically weighted by the
  inverse of an estimated variance which is not straightforward in the setting
  here, since the aim is to compare different methodological approaches
  performed within one study. As a consequence, the variance of the difference
  measure between, e.g., Kaplan-Meier and incidence proportion is
  required. This difficulty has, e.g., also been faced (but not solved) by
  \cite{lacny2015kaplan,lacny2018kaplan}, and
  Section~\ref{sec:assessmentofdiff} will explain how to bootstrap these
  variances.

  The remainder of the paper is organized as follows.
  Section~\ref{sec:def:even} discusses in detail the definition of AEs and of
  competing events and will also briefly consider composite events. The latter
  will be included to investigate the impact of ignoring censoring without the
  complication of competing events. Section~\ref{sec:org} explains the
  organization of the data analyses within SAVVY and may serve as a template
  for future investigations of related questions. Here, an important aspect is
  that trial level analyses will be run at sponsors' sites, but meta-analyses
  will be run centrally by the academic project collaborators. The statistical
  methods on trial level are collected in Section~\ref{sec:statmethods}. This
  section, in particular, explains how SAVVY will quantify and account for different lengths
  of follow-up and makes a connection between the different methods of
  estimation. For instance, the incidence proportion will equal the
  Aalen-Johansen estimator evaluated at the largest observed time in the
  absence of censoring. Details of the meta-analyses to be performed are in
  Section~\ref{sec:metaanalysis}. This section, in particular, addresses the
  multitude of comparisons to be considered as well as assessment of bias and
  heterogeneity. A brief discussion is in Section~\ref{sec:disc}, also
  addressing the question of recurrent AEs, and software code is provided in
  the online supplement.

\section{Definition of events}\label{sec:def:even}

\subsection{Adverse events}
According to the Good Clinical Practice (GCP) guideline an AE is defined as ``any untoward medical occurrence in a patient or clinical investigation subject administered a pharmaceutical product and which does not necessarily have a causal relationship with this treatment'' \citep{gcp_ich}. In this meta-analytic study, the choice of AEs within the
selected clinical trials is left to the sponsor. These may be defined as AEs of special interest, as belonging to a specific 
Medical Dictionary for Regulatory Activities (MedDRA) system organ class or preferred term, as being severe according to a toxicity grading, as being related to the investigational product, as serious, or as a combination of these characteristics.

Often sponsors will select as AEs the adverse drug reactions presented in the core data sheet for a first submission for drug approval, and select the studies that supported the frequency derivation.

These choices may result in a range of frequency, from common AEs to rare ones. It is expected that differences between the methodological approaches will be less marked with very rare AEs or very common AEs. So, grouping of different rare AE types into one AE category would be permissible for studying methodological differences. Making use of this frequency range (from rare to more common per selected trial) allows to further investigate the impact of the frequency on the differences between statistical analysis methods. Ideally, AEs of different frequencies per trial should be chosen, e.g. around 30\%, around 10\%, and around 1\%.

The investigation is restricted to the analysis of the occurrence of the first AE of a specific
type and will not consider the analysis of recurrent events. However, the
  relevance of the present investigation for recurrent AEs will also be considered in Section~\ref{sec:disc}. 

\subsection{Competing events definition}\label{sec:CE}

Competing events are events that preclude the occurrence of the AE of interest. For instance, if in a
clinical trial the focus is on estimating the probability of headache, patients who die without prior headache report will never report headache. It is
obvious that death is a competing event with respect to the occurrence of headache. However, defining a general rule as to which events should be
treated as a competing event, without specific insight into the specific event of interest and study at
hand, is challenging.

As a rule of thumb, any event that both a) would be viewed from a patient
perspective as an event of his/her course of disease or treatment, and b)
would stop the recording of the AE of interest should be viewed as a competing
event. This situation would typically occur when a patient discontinues the
treatment due to another AE judged by the investigator as too severe to
continue treatment, or when the patient discontinues treatment or study due to
progression/lack of efficacy, and, as a consequence, the recording of AEs
ends. Hence, if end of follow-up for AEs, withdrawal of consent or
discontinuation is disease or treatment-related, this would be
handled as a competing event. See \cite{law_cook_2019} for related
considerations.

In contrast to a competing event, the time-to-event is censored if the patient
reaches the designated end of follow-up without having had the AE of interest 
or a competing event as defined above. This situation is present with
administrative censoring due to the regular end of the trial or the end of
follow-up for AEs due to the planned end of treatment (often end of treatment plus an additional fixed time interval, e.g., 30
 days) not triggered by the course of disease.

In the analysis, the different competing events will be combined into one composite competing event, as the aim is to compare different methods to quantify the risk of an interesting AE and not the risk of a competing event of a specific type. 

There is much debate among statisticians on the question which events should
be analyzed as competing events and which events should lead to censoring the
time-to-event. In practice, there is no discussion that death without the
previous occurrence of the interesting AE acts as a competing event with
respect to AE occurrence. The reason is that after death the AE can
definitely not occur any more, and in this sense ``death without prior AE'' is a
so-called ``hard'' competing event. However, the other events mentioned above as
loss to follow-up, withdrawal of consent or treatment discontinuation can be
regarded as so-called ``soft'' competing events in the sense that, thereafter,
the interesting AE in principle still could occur, but cannot be observed due
to end of follow-up. Discussions therefore arise around the question
whether to treat these soft competing events as censoring or as competing
event. One extra concern here is that, e.g., treatment discontinuation will likely
alter the AE hazard. See also \cite{unkel2018} on how these aspects connect to the current debate on estimands.

Therefore, two different approaches will be compared in this project. In a first approach (called ``all-events approach'' in the following), all competing events mentioned above will be combined and analyzed as a single competing event and only patients reaching the designated end-of-follow-up with neither former AE of interest nor competing event as defined above will contribute as censored observation. In a second approach (referred to as ``death-only approach'' in the following), only the hard competing event (i.e. death without prior AE of interest) will be analyzed as competing event, and the soft competing events will be analyzed as censored observations.

\subsection{Composite events}

Additionally to the statistical analyses of AEs considering competing events, further analyses encompassing both the AE of interest and the competing event as composite event will be performed, thereby addressing the composite estimand \citep{unkel2018,rufibach_2019}. The time to the composite event will be defined as time to the interesting AE or to the competing event whatever occurs first, and patients with neither the interesting AE nor the competing event will contribute a censored time-to-event. The rationale for the inclusion of this approach is to gauge the impact of using time-to-event methodology to account for varying follow-up times without the methodological complication of competing events. To this end, time-to-event analyses accounting for censoring will be compared to the traditionally used incidence proportion.

\section{Organization of the data analysis}\label{sec:org}

The SAVVY project group consists of the academic project collaborators who
planned the statistical analyses (in the following referred to as the ``analysis
center'') and the participating sponsors, who contribute randomized clinical trial data for analysis. In the
SAVVY project, the data analysis involves the following steps:
\begin{enumerate}
\item {\it Pre-registration}:
Confidential pre-registration of the clinical trials selected by the sponsors with the analysis center and allocation of a SAVVY trial identifier (ID) to registered trials by the analysis center

\item {\it Individual trial analysis}:
Analysis of the registered trials at sponsor's site using code provided by the analysis center and transfer of aggregated trial level results to the analysis center 

\item {\it Meta-analysis}:
Meta-analysis of trial level results at the analysis center
\end{enumerate}
In the following, these steps will be considered one by one in more detail.

\paragraph{Pre-registration}

The sponsors select the randomized clinical trials and the AEs they wish to enter into this project. In order to avoid selection bias, these trials have to be confidentially pre-registered with the analysis center before running the analyses. For the identification of the trials, a unique trial ID according to a publicly accessible trial registry, e.g. clinicaltrials.gov or the German Clinical Trials Register, has to be provided together with some characteristics of the trial and the selected AEs (see Table~\ref{tab:prereg}). The identification of the trials included will not be disclosed otherwise. In publications or presentations it will only be reported that studies have been identified to the analysis center. The analysis center will handle any information related to trial level data in a confidential manner. Also, the exchange of trial related information between the sponsor and the analysis center will take place in a secured manner following the sponsors' individual policies regarding the secure exchange of confidential information.

To register the trials and the AEs for the SAVVY project, a spreadsheet
is filled in by the sponsor containing one row per AE of interest with the main 
AE characteristics, e.g. seriousness, severity, MedDRA system organ class (SOC), MedDRA preferred
term (PT). If a sponsor does not want to provide a particular information or if the information is not relevant due to the particular grouping applied, ``NA'' for ``not applicable'' can be used. Table~\ref{tab:prereg} shows the
characteristics of the selected trials and AEs that will be captured.

\begin{table}[htb]
\begin{center}
\caption{Characteristics of the selected trials and AEs captured during pre-registration}\label{tab:prereg}
\begin{tabular}{ll}
\hline
Trial / AE characteristic & Explanation, possible entries\\\hline
Unique trial ID & Clinical trials registry number according to a publicly  \\
&\hspace{0.1cm}  accessible trial registry, e.g. clinicaltrials.gov, German \\
& \hspace{0.1cm} Clinical Trials Register\\
Indication  & Indication / therapeutic area investigated in the trial\\
Type of comparison & Active or placebo controlled\\
End of the trial & Year of last patient / last visit of the trial\\
 Maximum follow-up time for primary & Maximum follow-up time of the patients for recording the  \\
\hspace{0.1cm}  efficacy endpoint & \hspace{0.1cm} primary efficacy endpoint (in days)\\
AE ID & AE identifier, incremental AE numbering (from 1 to total \\
& \hspace{0.1cm} number of selected AEs) per trial \\
 Maximum follow-up time for AE & Maximum follow-up time of the patients for recording \\
& \hspace{0.1cm} the AE (in days)\\
Seriousness of AE & Specify if serious AE, any AE or NA\\
Severity of AE & Specify CTCAE toxicity grade (e.g. $\ge 3$), any toxicity\\
& \hspace{0.1cm} grade or NA\\
MedDRA system organ class (SOC) of AE & Specify SOC(s), any SOC or NA\\
MedDRA preferred term (PT) of AE & Specify PT(s), any PT or NA\\
Special interest AE & Specify special interest AE, any AE or NA\\
Hard competing events & Specify the type(s) of the hard competing event(s) as\\
& \hspace{0.1cm} defined in Section~\ref{sec:CE}, e.g., death\\
Soft competing events & Specify the type(s) of the soft competing event(s) as\\
& \hspace{0.1cm} defined in Section~\ref{sec:CE}, e.g., end of treatment, withdrawal\\
& \hspace{0.1cm} of consent, progression\\

\hline
\end{tabular}
\end{center}
\end{table}

After receipt of the registration sheet, a SAVVY trial ID will be allocated to
the trial by the analysis center. The SAVVY ID will be entered into the study
characteristics spreadsheet and returned to the sponsor.

\paragraph{Individual trial analysis}

The analyses of the individual clinical trials will be done at sponsor's site using SAS or R code provided by the analysis center. Therefore, it is not required to release any individual patient data to the analysis center. Only aggregated data, summarizing the results of the analyses, will be shared with the analysis center. The analysis center will not share the aggregated data of one sponsor with any other sponsor. A manual is provided by the analysis center to the sponsor describing what needs to be done after receipt of the SAVVY trial ID and the program code. As a prerequisite, at the sponsors' sites, the individual clinical trials data sets must be brought into a format which allows the application of the provided SAS or R code. The required data structure is simple, similar to that of a standard survival analysis, and shown in Table~\ref{tab:structure}.

\begin{table}[htb]
\begin{center}
\caption{Required data structure for application of program code}\label{tab:structure}
\begin{tabular}{ll}
\hline
Column& Description\\\hline
AE ID & Number from 1 up to the number of selected AEs within the trial\\
Patient ID & Trial specific unique patient ID\\
Treatment group ID & Identifying which treatment is experimental and which is control
\\
Time to event & In days\\
Type of event & 1=AE of interest\\
& 2=hard competing event\\
& 3=soft competing event\\
& 0=censored\\
\hline
\end{tabular}
\end{center}
\end{table}

For each trial one dataset is needed in which all AE specific data for the selected AEs are set below each other. The different AEs are distinguished by the AE ID, matching the AE ID given in the study characteristics table filled for trial pre-registration. The treatment group ID will be used by the SAS or R code but is not included in the results where experimental and control groups are coded as ``A'' and ``B'', respectively. Observations with missing data, negative event times or type of event not in $\{0,1,2,3\}$ are automatically excluded from the analyses.

The SAVVY trial ID is inserted in the SAS or R code. The program returns the aggregated data summarizing the results of the statistical analysis methods described in Section~\ref{sec:statmethods} and the results of some further descriptive analyses on the AE and the competing event, namely mean, median, minimum and maximum event time by type of event and overall, and the total numbers of AEs, competing events, and censored observations, each overall and per treatment group. The dataset containing all results is named automatically identical to the SAVVY trial ID and is sent to the analysis center for further processing.

\paragraph{Meta-analysis} 

Once the results of all registered trials are received, the analysis center performs the meta-analyses described in Section~\ref{sec:metaanalysis} of the estimated parameters described in Section~\ref{sec:statmethods}. The results of the meta-analyses will be presented and discussed within the SAVVY project group without identifying individual trials and sponsors.

\section{Statistical methods on trial level}\label{sec:statmethods}

\subsection{Quantifying length of follow-up}\label{toitoitau}

We will use two approaches to both quantify length of follow-up and to study
its impact on analysing the occurrence of AEs. The first approach is guided by
the implicit choice made when calculating incidence proportions. The second
approach is guided by the concern that overly small risk sets late in time may
lead to unstable probability estimators \citep{pocock2002survival}.

The first approach is to choose a time $\tau_{\rm A}$ as the largest observed
time (censored or AE or competing event) which was (if observed AE) or could
have been (if censored or competing event) an observed AE time in group
A. Time point $\tau_{\rm B}$ is defined analogously for group~B. Then one step
to account for different lengths of follow-up between groups is to restrict
statistical inference to the smaller of the two time points. Hence, let $\tau
= \min(\tau_{\rm A}, \tau_{\rm B})$. The motivation behind this approach is
two-fold: Firstly, the commonly used incidence proportions are calculated in
the complete data set, i.e., for the data available on $[0, \tau_{\rm A}]$ and
$[0, \tau_{\rm B}]$, respectively. Secondly, group comparisons for time-to-event
data are typically restricted to the smaller of these two time intervals
only. For instance, the common log-rank test only compares groups as long as
the risk sets are non-empty in both groups.

The second approach follows a suggestion of \cite{pocock2002survival}. It covers a range of choices for quantifying length
of follow-up and will depend on the proportion of patients still at risk. So,
additionally, choose $\tilde\tau_{\rm A}(p)=\tilde\tau_{\rm A}$ as the time
such that $100\cdot p$\% of all patients in group~A are still at risk just
prior time~$\tilde\tau_{\rm A}$ and may have an observed event at
time~$\tilde\tau_{\rm A}$, $p \in (0, 1)$. To be precise, let $\tilde\tau_{\rm
  A}$ be the $100\cdot p$\%-quantile of the usual empirical distribution
function~$\hat F$ of the observed times (irrespective of censoring status),
\begin{displaymath}
  \tilde\tau_{\rm A} = \inf\{ t:\, \hat F(t)\ge p\}.
\end{displaymath}
Choose $\tilde\tau_{\rm B}$ analogously in group B and let $\tilde\tau =
\min(\tilde\tau_{\rm A}, \tilde\tau_{\rm B})$. We will consider $p\in\{0.3,
0.6, 0.9\}$. The relationship between $\tilde\tau_{\rm A}$ and $\tau_{\rm A}$
is that both time points coincide for the choice of $p=1$.

We also note that the different choices of $\tau$ account for different
time horizons underlying the estimation methods within groups, but they do not
account for differential drop-out between groups. It is, e.g., possible that
$\tau_{\rm A} = \tau_{\rm B}$, but that drop-out rates differ between
groups. Such differential drop-out would therefore be potentially treatment
group-related, suggesting to handle drop-outs as competing risks, see
Section~\ref{sec:CE}.

\subsection{One-sample estimators} \label{sec_one} Methods are exemplarily
discussed for group A and for $\tau$. The estimators of ``absolute AE risk''
that we will consider fall into three groups \citep{allignol2016}. Firstly,
the incidence proportion accounts for competing risks but not for censoring
(Equation~(\ref{eq:2}) below). Secondly, one minus the Kaplan-Meier estimator
accounts for censoring but not for competing risks (Equation~(\ref{eq:5})),
and this is also true for a standard conversion of the incidence density to a
probability (Equation~(\ref{eq:99})). Thirdly, the Aalen-Johansen estimator
generalizes the Kaplan-Meier estimator to competing risks and will later serve
as a benchmark or method of choice for nonparametric estimation of the
cumulative AE probability. The connection to the incidence proportion is that
both coincide in the absence of censoring. The connection to one
  minus the Kaplan-Meier estimator is that both coincide in the absence of
  competing risks. A parametric counterpart (Equation~(\ref{eq:7})) of the
Aalen-Johansen estimator (Equation~(\ref{eq:9})) based on incidence densities
is also considered. The incidence proportion divides the number
  of patients with an observed AE on $[0, \tau]$ in group A by the
  number~$n_{\rm A}$ of patients in group~A. More precisely, introduce individual first-AE-counting processes
  \begin{displaymath}
    N_i(t) \in \{0,1\},\, i\in\{1, \ldots, n_{\rm A}\},\, t\in [0, \tau],\ N_i(0)=0,
  \end{displaymath}
  where $N_i(t)=1$ denotes that an AE has been observed for patient~$i$ in the
  time interval~$[0,t]$ and that no competing event has been observed before
  the AE. Analogously, let
  \begin{displaymath}
    \bar{N}_i(t)\in \{0,1\},\, i\in\{1, \ldots, n_{\rm A}\},\, t\in [0, \tau],\ \bar{N}_i(0)=0,
  \end{displaymath}
  denote~$i$'s counting process of observed competing events. Because we
  consider time-to-first-event and type-of-first-event, we have that
  \begin{displaymath}
    N_i(t) + \bar{N}_i(t) \le 1
  \end{displaymath}
  and both~$N_i(t)$ and~$\bar{N}_i(t)$ change their value from~$0$ to~$1$ at
  most once, when a time-to-first-event has been observed. The aggregated
  processes are
  \begin{displaymath}
    N_{\rm A}(t) = \sum_{i=1}^{n_{\rm A}} N_i(t),\, \bar{N}_{\rm A}(t) = \sum_{i=1}^{n_{\rm A}} \bar{N}_i(t).
  \end{displaymath}
  In the absence of censoring, the sum of the two aggregated processes will
  eventually be equal to~$n_{\rm A}$, but in general we have $N_{\rm
    A}(\infty) + \bar{N}_{\rm A}(\infty) \le n_{\rm A}$. The incidence
  proportion now is
  \begin{equation}
    \label{eq:2}
    IP_{\rm A}(\tau) = \frac{N_{\rm
        A}(\tau)}{n_{\rm A}}.
  \end{equation}
  The incidence density has the same numerator, but divides by
  person-time-at-risk. Again, to be precise, introduce individual
  at-risk-processes
  \begin{displaymath}
    Y_i(t) \in \{0,1\},\, i\in\{1, \ldots, n_{\rm A}\},\, t\in [0, \tau],\ Y_i(0)=1,
  \end{displaymath}
  where, for~$t>0$, $Y_i(t)=1$ denotes that the patient is still under
  observation on~$[0,t)$ and that neither an AE nor a competing event have
  happend on~$[0,t)$. Note that the at-risk-processes are left-continuous,
  such that $Y_i(t)$ denotes the at-risk status just prior time~$t$. If
  $Y_i(t)=1$, an event may happen and be observed at time~$t$. Otherwise,
  $Y_i(t)=0$. The incidence density now is
  \begin{equation}
    \label{eq:4}
    ID_{\rm A}(\tau) = \frac{N_{\rm  A}(\tau)}{\int_0^\tau \sum_{i=1}^{n_{\rm A}} Y_i(u)\,{\rm  d}u}.
  \end{equation}
The incidence density is not a probability, but estimates the AE hazard with
values in~$[0, \infty)$ under a constant hazard assumption. A typical
transformation of this estimator onto the probability scale is
\begin{equation}
  \label{eq:99}
  1 - \exp\left(- ID_{\rm A}(\tau) \cdot \tau\right).
\end{equation}
Assuming a constant AE hazard, estimator~\eqref{eq:99} estimates the same
quantity as one minus the Kaplan-Meier estimator, which only
  codes observed AEs as an event and censors anything else. To be precise,
  introduce increments
  \begin{displaymath}
    \Delta N_i(t) = N_i(t) - \lim_{u \nearrow t} N_i(u),
  \end{displaymath}
  which equal one, if an AE (before any competing event) is observed for
  patient~$i$ at time~$t$, and $\Delta N_i(t) = 0$ otherwise. Defining $\Delta
  \bar{N}_i(t)$ analogously, the increments of the aggregated processes are
  \begin{displaymath}
    \Delta N_{\rm A}(t) = \sum_{i=1}^{n_{\rm A}} \Delta N_i(t),\, \Delta \bar{N}_{\rm A}(t) = \sum_{i=1}^{n_{\rm A}} \Delta \bar{N}_i(t).
  \end{displaymath}
  The size of the risk set is
  \begin{displaymath}
    Y_{\rm A}(t) = \sum_{i=1}^{n_{\rm A}}  Y_i(t),
  \end{displaymath}
  and one minus the Kaplan-Meier estimator which only codes observed AEs as an
  event and censors anything else can be expressed as
  \begin{equation}
    \label{eq:5}
    1 - \hat S_{\rm A}(\tau) = 1 - \prod_{u \in (0, \tau]}\left( 1 - \frac{\Delta N_{\rm A}(u)}{Y_{\rm A}(u)}\right) = 1 - \prod_{u \in (0, \tau]}\left( 1 - \Delta \hat{\Lambda}_{\rm A}(u)\right).
  \end{equation}
  Here, $\Delta \hat{\Lambda}_{\rm A}(u)$ is the increment of the
  nonparametric Nelson-Aalen estimator of the cumulative AE hazard
  \begin{equation}
    \label{eq:na}
    \hat{\Lambda}_{\rm A} (\tau) = \sum_{u \in (0, \tau]}\frac{\Delta N_{\rm A}(u)}{Y_{\rm A}(u)},
  \end{equation}
  where the product in~\eqref{eq:5} and the sum in~\eqref{eq:na} is over all
  observed, unique event times~$u$. Also note that we are slightly abusing
  notation in~\eqref{eq:5}, because~$\hat S_{\rm A}(\tau)$ is not estimating a
  proper survival function because of the presence of competing
    risks. The Nelson-Aalen estimator, however, is a proper estimator of the
  cumulative AE hazard. Assuming a constant AE hazard, $\hat{\Lambda}_{\rm A} (\tau)$ and $ID_{\rm
  A}(\tau) \cdot \tau$ estimate the same quantity.

Accounting for competing risks now requires to acknowledge that there also is
a competing hazard. To begin, we introduce a competing incidence density
\begin{equation}
  \label{eq:66}
  \overline{ID}_{\rm A} (\tau)= \frac{\bar{N}_{\rm  A}(\tau)}{\int_0^\tau \sum_{i=1}^{n_{\rm A}} Y_i(u)\,{\rm  d}u}.
\end{equation}
Also using $ID_{\rm A}(\tau)$ as defined above, we obtain an estimator of the
cumulative AE probability based on incidence densities and accounting for
competing risks,
\begin{equation}
  \label{eq:7}
  \hat p_{ID; {\rm A}} (\tau)= \frac{ID_{\rm A}(\tau)}{ID_{\rm A}(\tau) + \overline{ID}_{\rm
      A}(\tau)} \left(1-\exp(-\tau\cdot[ID_{\rm A}(\tau) + \overline{ID}_{\rm
      A}(\tau)])\right).
\end{equation}
The connection of this estimator to the incidence proportion is that the
leading factor on the right hand side of the previous display equals $IP_{\rm
  A}(\tau)$ in the absence of censoring and if $\tau = \tau_{\rm A}$
\citep{beyersmann2017florence}. In words, both of these quantities estimate
the anytime-AE-probability in this situation. In the presence of censoring,
quantity~\eqref{eq:7} estimates the cumulative AE probability assuming all
hazards to be constant. The nonparametric counterpart is the Aalen-Johansen
estimator of the so-called cumulative incidence function,
\begin{equation}
  \label{eq:9}
  CIF_{\rm A}(\tau) = 
  \sum_{u \in (0, \tau]} 
  \prod_{v \in (0, u)}\left( 1 - \Delta \hat \Lambda_{\rm A}(v)- \Delta
    \hat{\overline{\Lambda}}_{\rm A}(v)\right)\Delta
  \hat \Lambda_{\rm A}(u),
\end{equation}
where $\Delta \hat{\overline{\Lambda}}_{\rm A}(v)$ now is the increment of the
competing Nelson-Aalen estimator in analogy to $\overline{ID}_{\rm A}$. Note
that we have again slightly abused notation, writing $CIF_{\rm A}(\tau)$,
although this quantity is an estimator.

%%{\color{red}Letzten Abschnitt gel\"oscht/auskommentiert nach Claudias Anmerkung}
%
%Before turning to two-sample comparisons in the next subsection, we note that
%the SAVVY study will also investigate all-encompassing composite time-to-event
%analyses. The rationale is to gauge the impact of using time-to-event
%methodology to account for censoring/varying follow-up times without the
%methodological complication of competing risks. To this end, secondary
%analyses shall compare a one minus Kaplan-Meier curve analysis of the
%composite \emph{adverse event or competing event whatever occurs first} with
%the incidence proportion.

%  So far, the rationale has been that an analysis of adverse events
% should account for both censoring/varying follow-up times and competing
% risks. In order to gauge the impact of the former without the complication of
% competing risks, another secondary analysis shall compare a one minus
% Kaplan-Meier curve analysis of the composite \emph{adverse event or competing
%   event whatever occurs first} with the incidence proportion. These
% probability estimators shall be compared in analogy to
% Section~\ref{sec_2sample}, accounting for different time points as outlined in
% Section~\ref{sec_taus}. Note that the analysis is such that any event of
% type~1 or type~2 defined in Section~\ref{sec_cr} will be viewed as a composite
% event.

\subsection{Two-sample Comparisons} \label{sec_2sample}

In principle, many methods of two-sample comparisons are conceivable. We here
aim to consider one method that applies to all one-sample estimators and
provides a quantification of risk differences and relative risks, where
\textit{risk} here refers to a probability estimator as defined earlier. To this
end, assume that we have estimators
\begin{equation}
  \label{eq:14}
  \hat q_{\rm A}, \hat{\text{var}}(\hat q_{\rm A})=s^2_{\rm A}, \hat q_{\rm B}, \hat{\text{var}}(\hat q_{\rm B})=s^2_{\rm B},
\end{equation}
where $\hat q_{\rm A}$ and $\hat q_{\rm B}$ are probability estimators within
groups. That is, we have one line of values~(\ref{eq:14}) for each time point
for the incidence proportion, one line of values for each time point for
incidence densities (transformed onto probability scale) etc., for each
estimation method discussed in Section~\ref{sec_one} and for each evaluation
time point defined in Section~\ref{toitoitau}. Then, we estimate the risk
difference by
\begin{equation}
  \label{eq:15}
  \hat{RD} = \hat q_{\rm A} - \hat q_{\rm B}
\end{equation}
with
\begin{equation}
  \label{eq:16}
  s^2 = \hat{\text{var}}\left(\hat{RD}\right) = \hat{\text{var}}(\hat q_{\rm A}) + \hat{\text{var}}(\hat q_{\rm B})
\end{equation}
and approximate 95\% confidence interval
\begin{equation}
  \label{eq:17}
  \hat{RD} \pm z_{0.975}\cdot s,
\end{equation}
where~$z_{0.975}$ is the $0.975$ quantile of a standard normal distribution.

Relative risks are estimated by
\begin{equation}
  \label{eq:18}
  \hat{RR} = \frac{\hat q_{\rm A}}{\hat q_{\rm B}},
\end{equation}
and we base variance estimation and construction of approximate confidence
intervals on a log-transformation. So, e.g., using the delta method
\begin{equation}
  \label{eq:19}
  \hat{\text{var}}\left(\log \hat q_{\rm A}\right) = \left(\frac{1}{\hat q_{\rm
        A}}\right)^2 \hat{\text{var}}(\hat q_{\rm A}) = \hat{\sigma}^2_{\rm A}
\end{equation}
and
\begin{equation}
  \label{eq:20}
  \hat{\text{var}}\left(\log \hat q_{\rm A} - \log \hat q_{\rm B}\right) =
 \hat\sigma^2_{\rm A} +\hat\sigma^2_{\rm B} = \hat\sigma^2.
\end{equation}
This leads to the backtransformed approximate 95\% confidence interval
\begin{equation}
  \label{eq:21}
 \hat{RR}\cdot\exp\left(\pm z_{0.975}\cdot \hat\sigma\right).
\end{equation}

The primary comparison of methods will be based on probability estimators as
explained. However, because of the omnipresence of hazard ratios for
group comparisons, we will also consider comparisons on the hazard scale as
detailed below:
\begin{itemize}
\item An estimated hazard ratio (output from standard Cox software) only
  coding AEs as ``observed event'', together with an estimator of its
  variance and a confidence interval for the hazard ratio.

\item Ditto for the competing event, now only coding competing events as
  ``observed event''. One rationale here is to check whether
  relevant signals on the hazard scale would have been missed by
  ignoring competing risks. We reiterate that, as for all competing event
  analyses, this will be done with the hard and the soft definition given in
  Section~\ref{sec:CE}.

\item Ratios of incidence densities for AEs, with variance
  estimation analogous to above. The rationale here is to check whether the
  simple constant hazard framework, although potentially misspecified, leads
  to a reasonable approximation of the hazard ratio estimated in
  semi-parametric fashion.

\item Ratios of incidence densities for competing events.

\item Ratios of Nelson-Aalen estimators for AEs, with variance
  estimation analogous to above.  The rationale here is to compare the usual
  hazard ratio estimator not only with a very simple parametric counterpart,
  but also with a fully non-parametric competitor. Under a proportional
  hazards assumption, the ratio of Nelson-Aalen estimators also estimates the
  hazard ratio, but not under non-proportional hazards \citep{Ande:comp:1983}.
 \item  Ditto for the competing event.
\end{itemize} 

\subsection{Assessment of differences of estimators}\label{sec:assessmentofdiff}
The estimators in~\eqref{eq:14} and the derived information on estimated risk
differences and risk ratios are of standard input form for a
meta-analysis. However, the aim is a methodological comparison of different
methods for quantifying AE risk when applied to the very same data. To this
end, the information on variances so far does not suffice, but what we need is
an estimator of the variance between, e.g., the incidence proportion and the
Aalen-Johansen estimator when calculated on the same data set. As is obvious
from the formulae in Subsection~\ref{sec_one}, such estimators will in general
be dependent. Closed form variance estimators might be obtained using the
\textit{functional} delta method \citep{Gill:Joha:surv:1990}, but one would need
to derive and implement such estimators for every single methodological
comparison. We have therefore decided to follow the advice of \cite{ABGK} and
to resort to bootstrap variances, drawing with replacement from the individual
patients under an {\it i.i.d.}\ set-up. We also note that in a meta-analysis of
published data on the overestimation of the cumulative revision arthroplasty
using a Kaplan-Meier-type estimator, \cite{lacny2015kaplan} used
common approximations of the estimated variance of the hazard ratio for this
purpose. However, this approach estimates a different variance as the correlation structure is not accounted for.

\subsection{Implementation}
The estimators displayed in the Sections~\ref{sec_one} and \ref{sec_2sample}
are readily available in the statistical software SAS (SAS Institute, Cary,
NC, US) and R \citep{R}. The implementation of the estimators of the incidence
proportion and the two estimators based on the incidence density is
straightforward by the use of the formulae. In SAS software the \texttt{proc
  lifetest} calculates the one minus Kaplan-Meier estimator. In R it can
either be obtained by the \texttt{survfit} function of the \texttt{survival}
package \citep{therneau2000} or, as it is a special case of a
competing risk setting, the \texttt{etm} function of the
identically named package \citep{allignol2011} can also be used to calculate
both one minus the Kaplan-Meier estimator and the Aalen-Johansen
estimator. Depending on which SAS version is used the Aalen-Johansen estimator
can, on the one hand, be computed by the predefined \texttt{\%CIF} Macro. On
the other hand, in newer versions of SAS software the \texttt{proc lifetest}
specifying the event of interest in the option \texttt{failcode} can be used.

The first part of the two-sample comparisons, the risk
differences and relative risks with corresponding variances, can
be directly calculated by implementing the formulae. The Cox model and
therefore the estimated hazard ratio may be obtained by the use of the
\texttt{proc phreg} in SAS and with the function \texttt{coxph} in
R. In oder to estimate (event-specific) hazard ratios, e.g., for
  AE, a well known coding method is to also code observed competing events as
  ``censored''. A brief look at the simple incidence densities illustrates
  correctness of this method for analyzing hazards. The estimator of the
  cumulative AE probability in~\eqref{eq:7} demonstrates that all hazards
  enter probability calculations and hence, the ``code as censored'' approach is
  only available on the hazard level. In both of the software,
it can be easily switched which event is of interest, such that the
hazard analysis of the competing event can as easily be
obtained. The ratios of the incidence densities for the adverse as well as for
the competing event are easily calculated once the incidence density for the
one-sample estimators have been saved. The \texttt{proc lifetest} with the
\texttt{nelson} option gives the Nelson-Aalen estimator. Moreover, the
\texttt{mvna} function of the \texttt{mvna} package \citep{allignol2008}
returns this estimator in R.

SAS macro code and the corresponding function in R is available as supplementary material. The main macro code has
been written in SAS 9.4 software and checked in R by one of the authors (RS). It has subsequently
been checked in a small scale pilot study (VJ, KR, CS).

\section{Meta-analysis}\label{sec:metaanalysis}

Once the trial level data have been analysed using the methods described in Section~\ref{sec:statmethods}, results will be summarized across trials using the approaches listed below. Whereas individual trial data analyses will be run within the sponsor company, meta-analyses will be run on the calculated probability and hazard (ratio) estimates centrally at the analysis center, i.e. the institutions of the academic project group members. 

\subsection{Method comparisons} \label{sec_metacomp}

Table~\ref{tab:comparison} gives an overview of the planned method comparisons. We will distinguish between the two types of the competing events introduced in Section~\ref{sec:CE}, i.e., all comparisons will be performed for both types of competing events (death-only and all-events). Thereby, the main interest is in the `all-events' competing event. Especially, the comparison of Aalen-Johansen estimators based on the different competing event definitions is of interest. The two Aalen-Johansen estimators will be compared for the AE as well as for the competing event. Moreover, the comparisons in terms of hazard ratios will be conducted for the AE and for the competing event. All comparisons are conducted at the five follow-up times as defined in Section~\ref{toitoitau} and not only at the final time-point.
  
\begin{table}[htb]
\begin{center}
\caption{Planned comparisons. The quantities marked with $\star$ are calculated in both groups.}\label{tab:comparison}
\begin{tabular}{lll}
\hline
Target quantity & Benchmark & compared against \\\hline
AE probability $\star$ & Aalen-Johansen & Incidence Proportion\\
 & Aalen-Johansen & Probability Transform Incidence Density\\
 & Aalen-Johansen & 1-Kaplan-Meier\\
 & Aalen-Johansen & Probability Transform Incidence  \\
 & &  \hspace{0.1cm} Density accounting for competing events \\

Composite endpoint $\star$ & 1-Kaplan-Meier & Incidence Proportion\\
Hazard Ratio  & Cox & Ratio Incidence densities\\
& Cox & Ratio Nelson-Aalen estimators\\

\hline
\end{tabular}
\end{center}
\end{table}

\subsection{Assessment of bias}\label{bias}
The bias of the estimators will be assessed by comparison with
the benchmark estimator. This will be assessed visually using Bland-Altman plots
of the AE probability, the risk difference and the (log) relative
risks \citep{blandaltman}. As we consider a comparison of an estimator to a
benchmark, the benchmark estimator is plotted on the x-axis instead
of the mean \citep{krouwer2008}. With these plots, both the one-sample (AE
probability) as well as the two-sample (risk difference, relative risk)
situation can be considered.

\subsection{Frequency categories}\label{freqcat}
For the one-sample estimators, the possible change in frequency categories depending on estimation method will also be investigated. According to the European Commission's guideline
on summary of product characteristics (SmPC) \citep{smpc} the frequency categories are respectively classified as `very rare', `rare', `uncommon', `common' and `very common' when found to be
$<0.01\%, <0.1\%, < 1\%, <10\%, \ge 10 \%$. Frequency categories obtained with the estimators will be compared to frequency categories obtained with the benchmark estimator, i.e., the Aalen-Johansen estimator.

The comparison of the conclusions about the therapies' safety derived from the two-sample comparisons of the various approaches shall be compared in terms of statistical significance, clinical relevance and benefit assessment criteria \citep{iqwigmethodenpaper,kieser2005} against the Aalen-Johansen approach as benchmark in frequency tables.

\subsection{Assessment of precision}\label{precision}
As assessment of precision, the standard errors or the width of the confidence intervals of the estimators will be compared to the benchmark ones. This is done in terms of plots of the ratios of the standard errors for the methods with at most small to moderate bias. The consideration of precision is deemed useful only in the absence of any substantial bias. 

\subsection{Random effects meta-analysis and meta-regression}
A more formal assessment of difference between estimators and possible factors influencing these is carried out in form of random effects meta-analyses and meta-regressions.
These will model the ratios of
the estimators considered in Section~\ref{sec_metacomp} (i.e., respective estimator divided by benchmark). The standard errors of these
ratios will be needed for the meta-analysis. As noted in
Section~\ref{sec:assessmentofdiff}, the derivation of these standard errors is
complicated by the dependence of the estimators. Therefore, they will be obtained with a
bootstrap, see Section~\ref{sec:assessmentofdiff}. 

To be more precise, the estimator of the log-ratio
($\log(\text{estimator}/\text{benchmark})$) $\hat{\theta}_k$ is observed with
bootstrapped variance $\hat{\sigma}_k^2$ for each adverse event
$k=1,...,K$. Then a normal-normal hierarchical model (NNHM) \citep{hedgesolkin}
of the form $\hat{\theta}_k\vert\theta_k\sim N(\theta_k,\sigma_k^2)$ with
$\theta_k\vert\theta,\rho \sim N(\theta,\rho^2), k=1,...,K$, is
fitted, with $\rho^2$ denoting the between AE heterogeneity. Thereby, between adverse events variability is
introduced via $\theta_k=\theta+\epsilon_k$ with $\epsilon_k\sim
N(0,\rho^2)$. As the main interest is in the mean parameter $\theta$ the
marginal model $\hat{\theta}_k\vert \theta\sim N(\theta,\sigma_k^2+\rho^2),
k=1,...,K$, will be used. The between adverse
event variability~$\rho^2$ is estimated by the Paule-Mandel estimator as recommended by \citet{veroniki}.
As we are also interested in exploring any heterogeneity identified, possible 
sources of heterogeneity are assessed using meta-regression models
including, for example, the frequency of the adverse event and competing event
recording over time.

The meta-analysis and meta-regression will first be performed on AE level and not on study level, i.e., AEs of the same study will be assumed to be independent. In a next step, as the structure of these data are more complex than in standard meta-analyses, potentially additional hierarchy levels will be considered. In the NNHM described above, it is assumed that it is sufficient to model the between AE heterogeneity. However, it might be necessary to consider in addition for instance any heterogeneity between studies or indications.  Therefore, random effects not only for AE but also for study or indication  are considered in subsequent analyses to explore whether additional hierarchy levels improve model fit.

\section{Discussion}\label{sec:disc}
We have presented the rationale and statistical concept of the
  empirical, meta-analytical SAVVY study which is presently ongoing. The study
  aims  to investigate the impact of commonly used methods to quantify AE incidence
  which fall short of accounting for both varying follow-up times and competing
  risks. 

  The study described in this paper considers time-to-first-AE
  only, but not recurrent AEs. The reasons are four-fold. Firstly, we kept in mind the ultimate goal of safety evaluation in drug development i.e. accurately informing the product label adverse drug reaction section by providing the most relevant frequency category for SmPC (Summary of Product Characteristics) or frequency for US PI (US prescribing information). Secondly, the incidence proportion is only meaningful as an estimator of absolute AE risk for first AEs, but not for recurrent ones. The incidence density could be computed for recurrent events in a meaningful way, but the assumption of a constant AE hazard would then be an even more restrictive parametric model \citep{windeler1995events}. 
	Thirdly, censoring, varying follow-up times and
  competing events will be no less important when AEs can be recurrent. For
  general recurrent events analyses, this has only very recently be
  re-emphasized by \cite{andersen2019modeling}. Fourthly, in a
  time-to-first-event analysis, the absolute AE risk or cumulative AE
  probability, non-parametrically estimated by Aalen-Johansen, is a natural
  target quantity or estimand. In a recurrent events setting, the options for
  statistical modelling become more complex, because intermediate AEs will, in
  general, impact the incidence of subsequent AEs. One consequence is that in
  the time-to-first-event setting, the absolute AE risk can be expressed via
  fully conditional intensities, while the question of whether to use fully
  conditional or rather marginal approaches becomes a more pressing question
  when AEs are recurrent, see again \cite{andersen2019modeling}. It is our intention that the SAVVY project
  will, in the future, also further investigate the analysis of recurrent AEs,
  and, to begin, such investigations shall be informed by the results of the
  study described in the present paper.

The current investigations within the SAVVY project focus on analyses of individual studies. In practice, these analyses would be integrated across trials. Particular problems arise when only a small number of trials is combined in a random effects meta-analysis \citep{bender2018} or the events are rare \citep{gunhan2019}. Furthermore, strategies for signal detection in safety analyses are also not considered here, but are subject to an ongoing research (see e.g. \citet{gould2018}).

\vspace{0.5cm}
\noindent {\bf{Conflict of Interest}}

\noindent {\it{VJ, KR and FL are employees of Novartis Pharma AG (Basel,
    Switzerland), F.\ Hoffmann-La Roche (Basel, Switzerland) and Pfizer
    Deutschland (Berlin, Germany), respectively. TF has received personal fees
    for consultancies (including data monitoring committees) from Bayer,
    Boehringer Ingelheim, Janssen, Novartis and Roche, all outside the
    submitted work. JB has received personal fees for consultancy from Pfizer,
    all outside the submitted work. CS has received personal fees for
    consultancies (including data monitoring committees) from Novartis and
    Roche, all outside the submitted work. The companies mentioned will
    contribute data to the meta-analysis. RS has declared no conflict of
    interest. }}

\end{document}